\documentclass[twocolumn,showpacs,superscriptaddress,prl]{revtex4}

\usepackage{graphicx}                 

\usepackage{amsmath}                  
\usepackage{amssymb}

\def\beq{\begin{equation}}
\def\eeq{\end{equation}}
\def\bea{\begin{eqnarray}}
\def\eea{\end{eqnarray}}

\begin{document}

\preprint{in preparation for Phys. Rev. Lett.}

\title{X-ray scattering as a probe for warm dense mixtures
       and high-pressure miscibility}

\author{K.~W\"unsch}
\affiliation{Centre for Fusion, Space and Astrophysics,
             Department of Physics,
             University of Warwick,
             Coventry CV4 7AL, United Kingdom}
\author{J.~Vorberger}
\affiliation{Centre for Fusion, Space and Astrophysics,
             Department of Physics,
             University of Warwick,
             Coventry CV4 7AL, United Kingdom}
\author{G.~Gregori}
\affiliation{Clarendon Laboratory,
             University of Oxford,
             OX1 3PU, United Kingdom}
\author{D.O.~Gericke}
\affiliation{Centre for Fusion, Space and Astrophysics,
             Department of Physics,
             University of Warwick,
             Coventry CV4 7AL, United Kingdom}

\date{\today}                         

\begin{abstract}
We demonstrate the abilities of elastic x-ray scattering to yield
information on dense matter with multiple ion species and on the
microscopic mixing in dense materials. Based on partial structure
factors from {\em ab initio} simulations, a novel approach for the
elastic scattering feature is applied to dense hydrogen-beryllium
and hydrogen-helium mixtures. The scattering signal differs
significantly between single species, real microscopic mixtures,
and two separate fluids in the scattering volume.
\end{abstract}


\maketitle

Most materials in nature and technical applications consist of multiple atomic
species and their properties can be only understood if all mutual interactions
are considered. This is also true for high energy density science applied to
astrophysics and inertial confinement fusion. Indeed, objects such as old stars
and giant gas planets can be considered as a natural laboratory for the
equation of state of dense, high-pressure mixtures
\cite{DLFB_07,MHVT_08,NHKF_08,DG_09}. Demixing and the subsequent segregation
of the heavy elements can play a crucial role in the energy balance of these
objects \cite{G_99}. Moreover, the assumed separation into helium-rich and
helium-poor phases strongly influences the evolution and present internal
structure of gas planets and might be the reason for the observed low helium
concentration in the atmosphere and the high luminosity of Jupiter and Saturn.
Phase separation in hydrogen-helium mixtures has been predicted by quantum
simulation for conditions found in the interior of gas giants
\cite{LHR_09,WM_10}. However, the verification of these theoretical predictions
in an astrophysical setting is quite indirect and requires full scale planet
modeling where demixing or phase separation is one of many unknowns. Thus,
laboratory experiments are needed to investigate these extreme conditions. This
is even more true for such complex, but very interesting processes as the
formation of pure carbon from methane under the high pressures found in Neptune
\cite{ACST_97}.

Driven by the progress in inertial confinement fusion \cite{G_NIF,L_NIF},
mixing receives currently also much interest in laboratory settings. As the
boundary between the fuel and the ablator is hydrodynamically unstable
during the compression phase, fuel and shell material (beryllium or carbon)
mix. The degree of mixing strongly affects the performance of the target and
can even prevent sufficient burn or ignition at all \cite{Lindl,AM_book}.
However, experimental techniques to probe mixing under the extreme conditions
in fusion targets prove to be very difficult and only a few spectroscopic
investigations exist \cite{WSMK_07,WHMC_09}. These found however that present
models for the mixing during the compression underestimate the experimental
results.

In this Letter, we develop the theoretical framework for the application of
elastic x-ray scattering as a tool to investigate mixtures and the
high-pressure miscibility of materials. X-ray Thomson scattering has been shown
to robustly deliver basic plasma parameters like density, temperature, and ion
charge state as well as dynamic and structural properties of simple materials
\cite{GGLR_03,GLNL_07,GGGV_NP}. First experiments with two-component systems
like plastics (CH) or lithium hydrate (LiH) have been performed as well
\cite{GGCF_06,SRMI_07,KNCD_08,BKBB_09}. Hitherto the theoretical description
of the spectrum was however based on an ion structure that was generated from
a one-component calculation via \cite{GGCF_06}
\begin{equation}
S_{ab}(k) = \delta_{ab}
            + \frac{\sqrt{n_a n_b}}{n}\frac{Z_a Z_b}{\bar{Z}^2}
                                      \Big[ S_{ii}^{\rm 1C}(k) - 1 \Big] \,.
\label{S_from_1D}
\end{equation}
Here, $S_{ab}$ denote the partial structure factors. $n_a$ and $Z_a$ are the
density and the charge of the ions of species $a$. The structure for the
effective one-component system, i.e.\ $S_{ii}^{\rm 1C}$, is calculated using
an average ion charge state defined by 
$\bar{Z} \!=\! \sum_a n_a Z_a / \sum_a n_a$. Such a treatment is only exact in
the limit of weakly coupled plasmas that can be described by the random phase
approximation (RPA) \cite{T_81}. It is however unable to describe the highly
nonlinear effects in the structure of strongly coupled ions \cite{WHSG_08}.
Therefore, to apply x-ray scattering as a reliable diagnostic method for dense
matter requires a theoretical description that considers the full microscopic
structure in the materials tested including the nonlinear interplay between
different highly correlated ion species. In contrast to the approach of 
Ref.~\cite{GGCF_06}, the new approach should also allow for $Z$-dependent
screening clouds for different ion species.

Here, we derive a full multi-component description of the x-ray scattering
signal. The partial structure factors required are obtained via density
functional molecular dynamics simulations (details in
Refs.~\cite{VTMB_07,WVG_09}). We show that the elastic scattering feature is
very sensitive to the ratio of the different elements in the scattering volume.
Thus, it can be used as a probe for the degree of mixing in strongly compressed
samples. Moreover, we predict considerable differences in the scattering
strength from microscopically mixed systems and matter consisting of two phases.
The differences are particularly pronounced for hydrogen-helium mixtures under
conditions found in the interior of Jupiter. Thus, elastic x-ray scattering can
be used to investigate demixing of hydrogen and helium under planetary
conditions.

We briefly sketch our theoretical description of x-ray Thomson scattering that
is developed along the ideas of Chihara \cite{C_87,C_00}, but fully generalized
to plasmas with multiple ion species. The influence of the ions is however
quite indirect as x-rays are scattered by electron density fluctuations. The
scattered intensity is thus proportional to the total electron structure factor
$S_{ee}^{tot}(\mathbf{k},\omega)$, where $\omega$ and $k$ are the frequency and
wave number shifts of the photon, respectively \cite{GR_09}. The structure
factor can be expressed as the Fourier-transform of the intermediate scattering
function with respect to time. The latter is the correlation function of
electron densities
\beq
F_{ee}^{\rm tot} (\mathbf{k},t) =
                    \langle \varrho_e^{\rm tot}(\mathbf{k},t)
                            \varrho_e^{\rm tot}(-\mathbf{k},0) \rangle \,.
\label{int_scatt}
\eeq
Now we split the total electron density into free electrons and contributions
of core electrons associated with $N$ different ion species:
$\varrho_e^{\rm tot}(\mathbf{k},t) \!=\! \sum_a^N \varrho_{a}^c(\mathbf{k},t)
                                         \!+\! \varrho^f(\mathbf{k},t)$.
Applying this decomposition in the quadratic form \eqref{int_scatt}, we obtain
$N^2$ bound-bound terms, $N$ bound-free terms and one free-free term. All of
these terms can be treated as in the case of just one ion species. Special care
must be however given to core electrons belonging to different ions species
$\sum_{a \neq b}\langle \varrho_{a}^c(\mathbf{k},t)
                               \varrho_{b}^c(-\mathbf{k},0)\rangle$ as it
defines a distinct ion-ion scattering function without core excitations.

Evaluating the different density correlations, we obtain for the total electron
structure factor (in the following, the $k$-dependence is dropped for
simplicity)
\bea
S_{ee}^{tot}(\omega) &\!\!=\!\!& 
                \bar{Z} S_{ee}(\omega)
              + 2 \sum_{a} \sqrt{\bar{Z} \, x_a} \, f_{a} \, S_{ea}(\omega)
\nonumber\\          &\!\!~\!\!&
              + \sum_{a}  Z_a^{c} \, x_a \int\! d \omega^\prime \;
                                  \tilde S_a^{ce}(\omega - \omega^\prime) \,
                                         S_a^S(\omega^\prime)
\nonumber\\          &\!\!~\!\!&
              + \sum_{a,b} \sqrt{x_a x_b} \, f_{a}f_{b} \, S_{ab}(\omega) \,.
\label{structurefactor_1}
\eea
Here, we introduced the concentrations $x_a \!=\! n_a / \sum_a n_a$. $Z_a^c$
is the number of core electrons bound to ions of the species $a$. The first term
describes correlations between two free electrons. The next term
accounts for free-bound correlations where $f_a$ is the atomic/ionic form
factor of bound states of component $a$. The second line contains
self-contributions, i.e., internal excitations and bound-free transitions.
Except the summation over species, it is unchanged from its usual form. The
last term describes correlations between two bound electrons.

The main problem using the expression \eqref{structurefactor_1} is the fact
that all partial structure factors are connected. This becomes particularly
clear when considering $S_{ee}$ which contains correlations between two
screening clouds and, thus, also ionic properties. Indeed, the structure
factors used in Eq.~\eqref{structurefactor_1} form a set of 
$\frac{1}{2} N (N \!+\! 1)$
equations \cite{blue}. This set can be conveniently written in matrix form and
then inverted. Such a procedure yields
\beq
\bar{Z} S_{ea}(\omega) = x_a \; q_a \, S_{aa}(\omega)
                        + \sum_{b \neq a} x_b \; q_b \, S_{ab}(\omega)
\eeq
for the free-bound structure factor. Here, $S_{ab}$ denotes the partial
ion-ion structure factors. The correlations of the free electrons to the ions
are contained in the screening function $q_a(k)$ which are defined via
\beq
q_a(k) = \frac{n_e C_{ea}(k) \chi^0_e(k)}{1 - n_e C_{ee}(k) \chi^0_e(k)} \,,
\label{screening}
\end{equation}
where $C_{ee}$ and $C_{ea}$ are the direct electron-electron and electron-ion
correlation functions, respectively. In lowest order, these are given by the
respective potentials. $\chi^0_e(k)$ denotes the density response of a free
electron gas \cite{blue}.

For the free-free structure factor, one obtains
\beq
\bar{Z} S_{ee}(\omega) = \sum_{a,b} \sqrt{x_a x_b} \, q_a q_b \, S_{ab}(\omega)
                         + S_{ee}^0(\omega) \,,
\eeq
which now separates electron-ion correlations from the structure factor of the
free electron gas $S_{ee}^0(k,\omega)$ that characterizes the kinetically free
electrons in the system.

The results can be summarized in the total electron structure factor of the
form
\bea
S_{ee}^{\rm tot}(\omega) &\!\!=\!\!&
              \sum_{a,b} \sqrt{x_a x_b} 
              \left[ f_a + q_a \right]\!\left[ f_b + q_b \right] S_{ab}(\omega)
              + \bar{Z} S_{ee}^0(\omega)
\nonumber \\             &\!\!~\!\!& 
              + \sum_a Z^c_a \,x_a \int d\omega^\prime \,
                \tilde S_a^{ce}(\omega-\omega^\prime) S^S_a(\omega^\prime) \,.
\label{sk_multi}
\eea
The first term describes quasi-elastic scattering at bound electrons and the
screening clouds associated to different ion species. In this contribution,
the full ionic structure, expressed by the partial structure factors $S_{ab}$,
influences the scattering spectrum. As the ion motion cannot be resolved in
current laser experiments, the ion structure can be treated statically:
$S_{ab}(k,\omega) \!=\! S_{ab}(k) \!\times\! \delta(\omega)$. The static
structure factors $S_{ab}(k)$ can be obtained by means of classical hypernetted
chain (HNC) calculations \cite{WHSG_08} or numerical simulations
\cite{HM_book,WVG_09}. The second term contains the full dynamic response of
the free electron gas which can be in lowest order described by the random
phase approximation. Extensions include electron-ion collisions and local field
corrections \cite{FWR_10}. The last term describes contributions due to the
the excitation or ionization of bound electrons by x-rays.

The inelastic contributions due to scattering at free electrons, bound-free
transitions, and internal excitations are unchanged from a description for
systems with one ion species \cite{GGCF_06}. We thus concentrate here on
elastic scattering, that is the first term of Eq.~\eqref{sk_multi}, which
highlights the mutual correlations between the different ion species. To 
calculate the weight of the Rayleigh peak, i.e.,
$W_R(k) \!=\! \sum_{a,b} \sqrt{x_a x_b} \left[ f_a(k) \!+\! q_a(k) \right]
              \left[ f_b(k) \!+\! q_b(k) \right] S_{ab}(k)$, we use structure
factors from {\em ab initio} simulations (DFT-MD) \cite{WVG_09} or solutions of
the hypernetted chain equations (HNC) \cite{WHSG_08}. The screening functions
are used in linear response to a Coulomb field, i.e., 
$q_a(k) \!=\! Z_a \kappa^2 / (\kappa^2 \!+\! k^2)$, where $\kappa$ is the
inverse electron screening length. The form factors $f_a$ are taken from
isolated atoms/ions \cite{DP_92}.

\begin{figure}[t]
\includegraphics[width=0.48\textwidth]{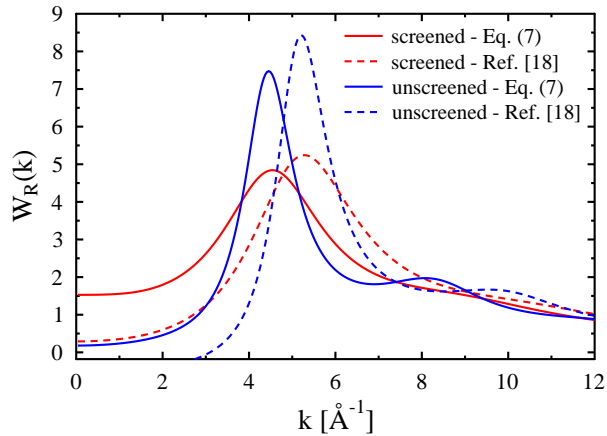}
\caption{Comparison of different treatments for the weight of the
         Rayleigh peak $W_R(k)$ (see text). Considered is a CH-plasma
         with $n_{\rm C} \!=\! n_{\rm H}
                    \!=\! 2.3 \!\times\! 10^{23}\,$cm$^{-3}$ and
         a temperature of $T \!=\! 1\,$eV. The ion charge states
         are $Z_{\rm C} \!=\! 4$ and $Z_{\rm H} \!=\! 1$. Both
         1C and partial structure factors are calculated via HNC
         \cite{WHSG_08}.}
\label{fig_Wk_ch}
\end{figure}

First we compare the full multi-component formula~\eqref{sk_multi} and the
approximate treatment of Ref.~\cite{GGCF_06} which is based on the structure of
an average ion component via Eq.~\eqref{S_from_1D}. Both approaches agree for
weakly coupled systems, but large differences occur for strongly coupled
systems as shown in Fig.~\ref{fig_Wk_ch} ($\Gamma_{ii} \!\approx\! 80$). For
both the screened and unscreened ion systems, the maximum of the Rayleigh
peak is shifted. This reflects the fact that the higher charged carbon
ions imprint their structure into the proton subsystem which cannot be
described by the one-component structure. Moreover, the mutual screening of the
ions is neglected in the reduced model which results in a strongly
underestimated Rayleigh peak at small $k$. The screened interactions are more
realistic \cite{GGGV_NP}, but the results for unscreened ions demonstrate that
the reduced model may even predict negative $W_R$ (especially, for small $k$).
For strongly coupled multi-component systems, the analysis should thus be based
on the new expression \eqref{sk_multi}.

Let us now turn to the application of the new multi-component description for
the mixing of beryllium and hydrogen as it occurs during inertial fusion
experiments. Due to the strong drive, the initially well-defined interface
between the two materials experiences a Rayleigh-Taylor instability. At this
stage, a volume element close to the original boundary will contain two fluids.
However, these fluids consist of either pure beryllium or hydrogen and the
system is made of two distinct phases. Later both materials will
microscopically mix due to diffusion. 

\begin{figure}[t]
\includegraphics[width=0.48\textwidth]{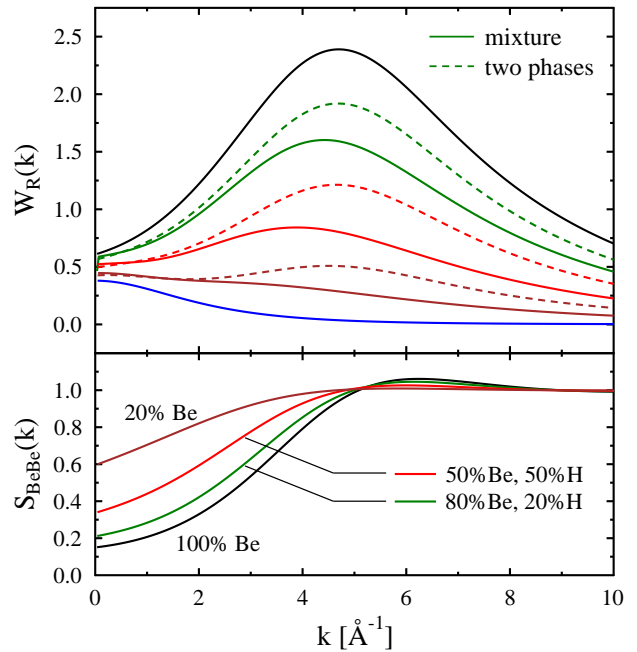}
\caption{(color online) The upper panel shows the weight of the
         Rayleigh peak, $W_R(k)$, for real microscopic mixtures of
         beryllium and hydrogen compared to systems that contain
         both materials in two pure phases. The concentrations
         of beryllium are (from top to bottom): 100\%, 80\%, 50\%,
         20\%, and 0\%. The temperature of the system is
         $T \!=\! 13\,$eV. The total densities were arranged to be
         in pressure equilibrium with pure beryllium at
         $n_{\rm Be} \!=\! 3.707 \!\times\! 10^{23}$\,cm$^{-3}$. 
         The lower panel displays the partial beryllium-beryllium
         structure factor for the same concentrations.}
\label{fig_BeH}
\end{figure}

Figure \ref{fig_BeH} demonstrates that the degree of mixing in dense
beryllium-hydrogen systems can be investigated by measuring the strength of
elastic x-ray scattering. To allow for experimental verification, we have
taken the density and the temperature from a recent experiment on shocked pure
beryllium \cite{LNCD_09}. Behind the shock front, all possible mixtures will be
compressed until they are in pressure equilibrium with the driven beryllium.
Thus, we have adjusted the total densities of the mixtures until the pressure
matches the one of the pure beryllium. This procedure yields, e.g., a density
of $n_{\rm H} \!=\! 8 \!\times\! 10^{23}\,$cm$^{-3}$ for the case of 100\%
hydrogen.

As beryllium scatters more efficiently than hydrogen, the elastic feature is
a strong function of the mixing ratio. Interestingly, both microscopic mixtures
and two-phase systems scatter very similar at small $k$. Here, the signal is
mainly determined by the partial beryllium-beryllium structure factor and we
can use these data to determine the concentrations. However, significant
differences arise at the maximum of the elastic scattering peak around
$k \!=\! 5\,$\AA$^{-1}$ that can be used to distinguish between hydrodynamic
and diffusive mixing.

\begin{figure}[t]
\includegraphics[width=0.48\textwidth]{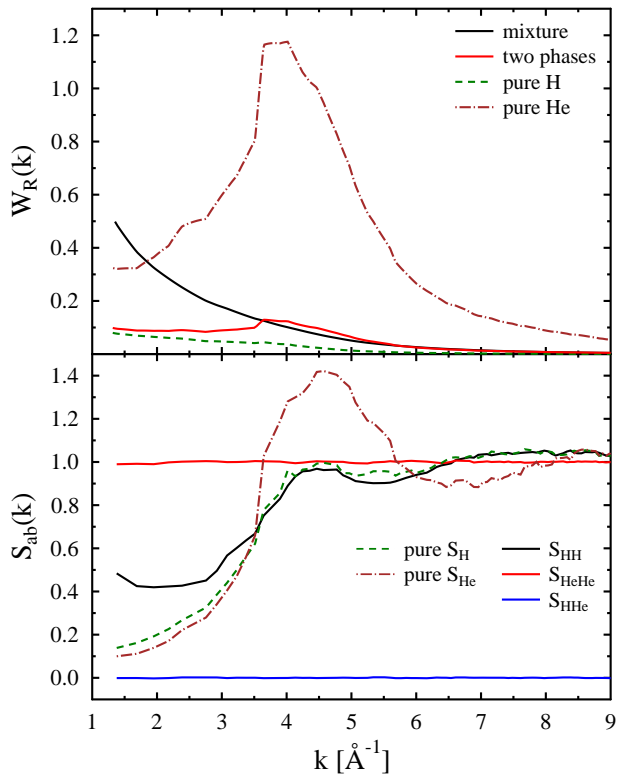}
\caption{(color online) Strength of the elastic scattering peak
         for pure hydrogen, pure helium, and a mixture with
         $x_{\rm H} \!=\! 0.924$ and $x_{\rm He} \!=\! 0.076$
         at a temperature of $T \!=\! 5000\,$K (upper panel).
         All systems have an electron density of
         $n_e \!=\! 4.7 \!\times\! 10^{23}\,$cm$^{-3}$.
         The lower panel shows the partial structure factors of
         the three systems above.}
\label{fig_HeH}
\end{figure}

In the next example, we investigate a hydrogen-helium mixture under conditions
as found in the interior of giant gas planets 
[$x_{\rm He} \!=\! n_{\rm He} / (n_{\rm He} \!+\! n_{\rm H}) \!=\! 0.0756$].
Density and temperature yield a system in the atomic/molecular
phase (no ionization). Fig.~\ref{fig_HeH} shows that the strength of the
elastic scattering feature displays large differences between the pure
substances and the mixture. Indeed, we find very distinct scattering features
for the different systems: pure helium scatters by far the most effective and 
displays a peak around $k \!=\! 4\,$\AA$^{-1}$, the mixture shows
a monotonically decreasing shape, and $W_R(k)$ for pure hydrogen is almost
featureless. Due to the mixing ratio, the two phase system is dominated by the
properties of hydrogen.

The reason for this behavior lies not only in the fact that helium atoms
scatter the x-rays more efficiently. The different structure factors play also
an important role. These structure factors were calculated from {\em ab initio}
quantum simulations (see Ref.~\cite{VTMB_07}) which allows us to treat neutral
systems as well. Pure helium has a strong peak in the structure factor which is
also clearly visible in the elastic scattering peak. This feature disappears in
the mixture as the helium concentration is low. However, the small fraction of
helium mitigates the correlations within the hydrogen subsystem. Thus, small
hydrogen-helium correlations lead to a shape of $W_R$ that is dominated by
the decreasing atomic form factor. Similar results are also found for denser
and hotter systems where hydrogen and helium are ionized (not shown).

In conclusion, we have derived a novel description for x-ray scattering in
systems with multiple atomic or ion species that take all the mutual
correlations into account. The results show that x-ray scattering is a powerful
tool to investigate dense mixtures and the mixing/demixing of materials under
extreme conditions as they are found during inertial confinement fusion and in
the interior of of old stars and giant planets.

\begin{acknowledgments}
We acknowledge financial support from EPSRC (via grants EP/D062837
and EP/G007187/1) and STFC.
\end{acknowledgments}


\end{document}